\documentclass[letter]{aa} 

\usepackage{graphicx}
\usepackage{txfonts}

\def\Herschel{\mbox{\it Herschel}}

\def\Nly{\mbox{$N_{\rm Ly}$}}
\def\Lbol{\mbox{$L_{\rm bol}$}}
\def\Vthick{\mbox{$V_{\rm thick}$}}
\def\Vthin{\mbox{$V_{\rm thin}$}}
\def\DVthin{\mbox{$\Delta V_{\rm thin}$}}
\def\dv{\mbox{$\delta V$}}

\def\HCOp{\mbox{HCO$^+$}}
\def\HCOpI{\mbox{H$^{13}$CO$^+$}}

\def\HII{H{\sc ii}}

\def\kms{\mbox{km~s$^{-1}$}}

\def\mic{\mbox{$\mu$m}}

\usepackage{amstext}

\begin{document} 

\title{Origin of the Lyman excess in early-type stars\thanks{Based on observations
carried out with the IRAM 30m Telescope. IRAM is supported by INSU/CNRS
(France), MPG (Germany) and IGN (Spain).}}

\author{
        R. Cesaroni \inst{1}
        \and \'A. S\'anchez-Monge \inst{2}
        \and M.T.~Beltr\'an \inst{1}
        \and S.~Molinari \inst{3}
        \and L.~Olmi \inst{1}
        \and S.~P.~Trevi\~no-Morales \inst{4}
}
\institute{
           INAF, Osservatorio Astrofisico di Arcetri, Largo E. Fermi 5,
           I-50125 Firenze, Italy
           \email{cesa@arcetri.astro.it}
\and
          I. Physikalisches Institut der Universit\"at zu K\"oln,
          Z\"ulpicher Strasse 77, 50937, K\"oln, Germany
\and
          INAF, Istituto di Astrofisica e Planetologia Spaziale,
          Via Fosso del Cavaliere 100, I-00133, Roma, Italy
\and
          IRAM,
          Avenida Divina Pastora, 7, N\'ucleo Central
          E-18012 Granada, Espa\~na
}
\offprints{\email{cesa@arcetri.astro.it}}
\date{Received date; accepted date}

\abstract{
Ionized regions around early-type stars are believed to be well-known objects,
but until recently, our knowledge of the relation between the free-free
radio emission and the IR emission has been observationally hindered by the limited angular resolution in the far-IR. The advent of \Herschel\ has now made it
possible to obtain a more precise comparison between the two regimes, and it
has been found that $\text{about a third}$ of the young \HII\ regions emit more Lyman
continuum photons than expected, thus presenting a {\it \textup{Lyman excess}}.
}{
With the present study we wish to distinguish between two scenarios that have been proposed
to explain the existence of the Lyman excess: (i)~underestimation of the
bolometric luminosity, or (ii)~additional emission of Lyman-continuum photons from an
accretion shock.
}{
We observed an outflow (SiO) and an infall (\HCOp) tracer toward a complete
sample of 200 \HII\ regions, 67 of which present the Lyman excess. Our goal
was to search for any systematic difference between sources with Lyman excess
and those without.
}{
While the outflow tracer does not reveal any significant difference between
the two subsamples of \HII\ regions, the infall tracer indicates that the
Lyman-excess sources are more associated with infall signposts than the other
objects.
}{
Our findings indicate that the most plausible explanation for the Lyman excess
is that in addition to the Lyman continuum emission from the early-type star, UV photons are emitted from accretion shocks in the stellar
neighborhood. This result suggests that high-mass stars and/or stellar
clusters containing young massive stars may continue to accrete
for a long time, even after the development of a compact \HII\ region.
}
%

\maketitle

%
\section{Introduction\label{sint}}

Early-type stars are well-known emitters of copious amounts of ultraviolet photons shortward of 912~\AA, which are sufficiently energetic to ionize atomic hydrogen.  The neutral gas surrounding such stars during their earliest stages is thus ionized by the Lyman-continuum photons, and an \HII\ region is created. In the ideal case of an optically thin \HII\ region around a single OB-type star, it is easy to obtain the stellar Lyman-continuum photon rate (\Nly) -- as well as other relevant parameters of the star -- from the radio continuum flux density emitted by the ionized gas (see, e.g., \citealt{scme}). In turn, from \Nly\ we can estimate the stellar luminosity (see, e.g., \citealt{pana, marti}). The latter can be measured by other means from the IR emission of the dusty envelope enshrouding the star and then compared to the value obtained from the Lyman continuum. While the two luminosity estimates should match, in practice the one obtained from the IR emission is often significantly greater than that computed from the radio flux. This discrepancy was discussed by \citet{wc89a}, and a number of explanations were proposed (opacity of the free-free emission, stellar multiplicity, insufficient angular resolution in the IR). The most important of these was probably the enormous difference in resolving power between IR and radio observations. In fact, the bolometric luminosity (\Lbol) estimate was based on the IRAS data, whose instrumental beam in the far-IR is $>$1\arcmin much greater than the typical size of a compact (i.e., young) \HII\ region ($<$10\arcsec; see, e.g., \citealt{wc89a}) imaged with a radio interferometer. Consequently, the IRAS fluxes are measured over a solid angle encompassing not only the \HII\ region, but also other (unrelated) stars, and the value of \Lbol\ is overestimated.

\begin{figure}
\centering
\resizebox{7.5cm}{!}{\includegraphics[angle=0]{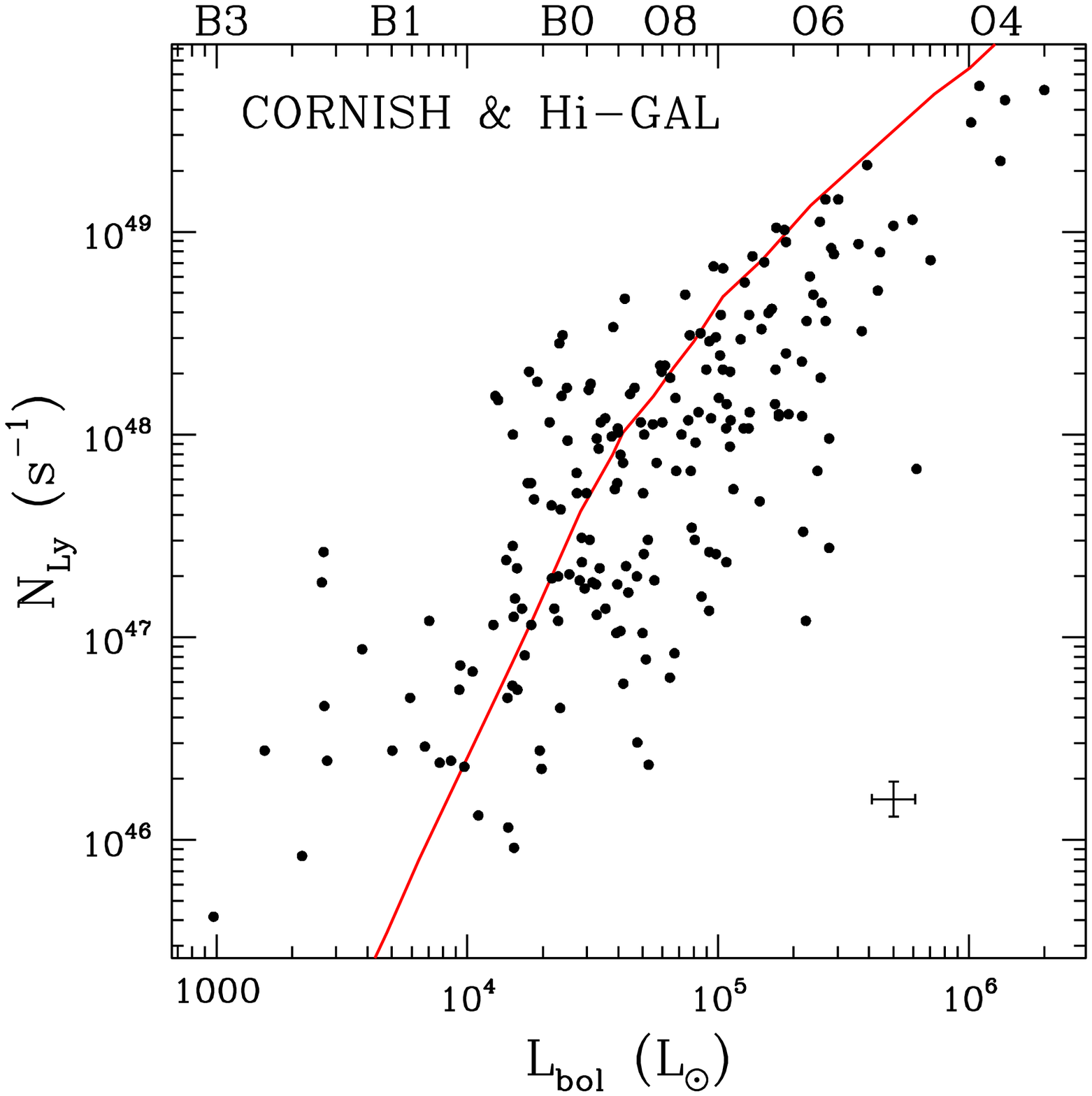}}
\caption{Plot of the Lyman-continuum photon rate versus the corresponding bolometric luminosity for the 200 CORNISH \HII\ regions with \Herschel/Hi-GAL counterparts studied by C2015. The spectral types indicated at the top are computed under the assumption that \Lbol\ originates from a single ZAMS star. The red solid line denotes the \Nly--\Lbol\ relation expected if the \HII\ regions were powered by single ZAMS stars (see C2015). Note that 67 objects ($\sim$34\% of the sample) lie above this line. The error bars in the bottom right corner correspond to a 20\% calibration uncertainty.}
\label{fpap7}
\end{figure}

With the advent of the ESA \Herschel\ Space Observatory \citep{pilb}, it was possible to dramatically improve on the \Lbol\ estimate and thus achieve a more reliable comparison with the luminosity obtained from the Lyman continuum. In a recent study, \citet{pap7} (hereafter C2015) have compared \Nly\ to \Lbol\ for the sample of compact and ultracompact \HII\ regions identified by \citet{purc} in the CORNISH survey \citep{hoare}. The value of \Lbol\ has been estimated for 200 objects by reconstructing the corresponding spectral energy distributions using the far-IR images of the \Herschel/Hi-GAL survey of the Galactic plane \citep{higal} and ancillary data from 1~mm to the mid-IR. As discussed by C2015 (see their Sect.~4.1), the surprising result is that about one-third of the \HII\ regions appear to have \Nly\ {\it \textup{greater}} than expected on the basis of their luminosities (see Fig.~\ref{fpap7}). After taking a number of possible explanations into account, C2015 (see also \citealt{sanch}) demonstrated that this so-called Lyman excess cannot be easily justified. Basically, two scenarios are possible.

The first involves the flashlight effect (e.g., \citealt{yobo}), where most stellar photons are leaking along the axis of a bipolar outflow, where the gas density is lower. In this way, a large portion of the photons is emitted away from our line of sight, and the assumption of spherical symmetry adopted in the \Lbol\ calculation leads to an underestimate.

The second scenario assumes that the Lyman excess is due to additional emission of UV photons from the stellar neighborhood. The models adopted to estimate \Nly\ for zero-age main-sequence (ZAMS) stars refer to visible objects, whose properties might be significantly different from those of a very young deeply embedded OB-type star, perhaps still undergoing accretion from the parental cloud. Recent theoretical studies of this type of objects predict significant Lyman-continuum emission from accretion shocks (\citealt{msmith, hosom}; Hosokawa pers. comm.), which may justify an excess of up to two orders of magnitude above the expected value of \Nly\ (see Figs.~15 and~16 of \citealt{msmith}).

If the first scenario is correct, molecular outflows should be more common in Lyman-excess sources than in the rest of the sample. In contrast, the second hypothesis implies that accretion should be preferentially present in the Lyman excess sample. With this in mind, we decided to search for outflow and infall tracers in the whole sample of 200 \HII\ regions studied by C2015 and thus compare the occurrence of these two phenomena in sources with and without Lyman excess. For this purpose, we observed the \HCOp(1--0) and SiO(2--1) lines. The former appears to present redshifted self-absorption when the gas is infalling, while the latter is a well-known tracer of shocked gas in molecular jets and outflows (see, e.g., \citealt{sep10, sep11}, and references therein).

%
\section{Observations and data analysis\label{sobs}}

The observations were performed with the IRAM 30 m telescope on Pico Veleta, using the EMIR receivers \citep{cart} with the fast Fourier Transform Spectrometer (FTS) at 200 kHz resolution \citep{klein}. By combining two bands of the EMIR receiver, we covered a bandwidth of 16 GHz in dual polarization. While a plethora of potentially useful lines was observed with our setup, in this study we focus on the \HCOp(1--0), \HCOpI(1--0), and SiO(2--1) rotational transitions. The observations were performed in December 2014 and May and August 2015 using the position-switch mode, where the reference position was chosen from the \Herschel/Hi-GAL maps as the closest to the target with minimal continuum emission at 350~\mic. In this way, we maximized the likelihood that molecular line emission is also very faint. The on-source integration time per target was 4~min, although for a limited number of faint emitters the integration was repeated twice. We pointed on strong nearby quasars every $\sim$1.5~hr. Pointing corrections were stable, with errors below 3". Typical system temperatures ranged from 105~K to 200~K at 3~mm and from 120~K to 400~K at 2~mm. The instrumental half-power beam width was $\sim$27\arcsec.

The data reduction and analysis were made with the program CLASS of the GILDAS package\footnote{The GILDAS software has been developed at IRAM and Observatoire de Grenoble -- see http://www.iram.fr/IRAMFR/GILDAS}. For each line all relevant spectra were averaged and a first-order polynomial baseline was subtracted. Then a Gaussian profile was fit to the \HCOpI(1--0) and SiO(2--1) lines. The \HCOp(1--0) line is often asymmetric, hence the fit took only the channels around the peak into account and ignored those where the asymmetry was most evident. In this way, we could obtain a reliable estimate of the peak velocity, which is the only useful parameter for our purposes.

%
\section{Results\label{sres}}

We obtained data for all of the 200 targets, 67 of which present Lyman excess. The detection rates for the different lines are summarized in Table~\ref{tdet}. The \HCOp(1--0) line was detected in all sources, whereas the \HCOpI(1--0) and SiO(2--1) transitions were detected in 192 and 99 targets, respectively. In the following we look for possible evidence in favor of either explanation of the Lyman excess phenomenon: flashlight effect due to a bipolar outflow, or accretion shock associated with infall.

\begin{table}
\caption[]{Number of targets detected in the different tracers and corresponding detection rates.}
\label{tdet}
\begin{tabular}{lccc}
\hline
\hline
 & with Lyman excess & w/o Lyman excess & total \\
\hline
\hline
all & 67 & 133 & 200 \\
\hline
SiO & 18 (27\%) & 81 (61\%) & 99 (50\%) \\
\hline
\HCOp & 67 (100\%) & 133 (100\%) & 200 (100\%) \\
\HCOpI & 63 (94\%) & 129 (97\%) & 192 (96\%) \\
infall$^a$ & 31/56=55\% & 41/115=36\% & 72/171=42\% \\ 
$E$$^b$ & $0.32\pm0.09$ & $0.10\pm0.06$ & $0.17\pm0.05$ \\
\hline
\end{tabular}

\vspace*{1mm}
$^a$~targets with $\delta V<-3\sigma$ (see text). Note that only 171 targets
could be used to estimate the \dv\ parameter. \\
$^b$~see Eq.~(\ref{eqe}) for the definition.
\end{table}

\subsection{Outflow}

If the flashlight effect is at work in the Lyman-excess sources, the number of outflows in these objects should outnumber those detected in the remaining targets. Since, as previously mentioned, SiO is an excellent tracer of shocks in jets and outflows, we would expect the SiO line to be found preferentially in the Lyman-excess sample. However, this is not the case, as only 18 targets out of 67 have been detected among the Lyman-excess sources (see Table~\ref{tdet}). The corresponding detection rate is 27$\pm$5\%, significantly lower than for the other \HII\ regions (61$\pm$4\%). This difference might be a luminosity effect. The Lyman-excess \HII\ regions are on average less luminous than the other sources (the former span the \Lbol\ range $10^3$--$1.7\times10^5~L_\odot$, the latter $10^4$--$1.3\times10^6~L_\odot$ -- see Fig.~\ref{fpap7}) and it may reasonably be expected that more intense SiO emission is associated with more luminous objects. There is no evidence that the SiO emission is preferentially associated with the Lyman-excess sources.

\begin{figure}
\centering
\resizebox{7.0cm}{!}{\includegraphics[angle=0]{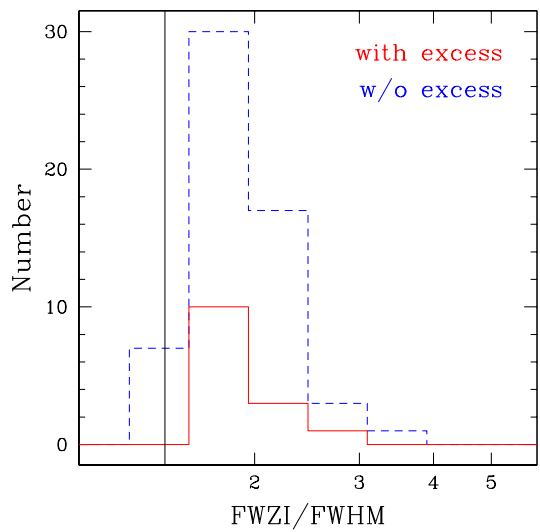}}
\caption{Distributions of the ratio FWZI/FWHM of the SiO(2--1) line for the sources with (red solid histogram) and without (blue dashed histogram) Lyman excess. The FWZI is measured at 25\% of the line peak (see text). The vertical line denotes the value of the ratio expected for a Gaussian profile. The bin size has been chosen following the Freedman-Diaconis rule.}
\label{ffws}
\end{figure}

Broad wings are indicative of outflow motions, and it is thus worth inspecting the full width at zero intensity (FWZI) of the SiO(2--1) line. In general, the FWZI is measured at the 3\,$\sigma$ level, but this definition affects the comparison between sources with different signal-to-noise ratios. Therefore, we prefer to define the FWZI as the line width at a fixed fraction of the peak intensity, which we arbitrarily chose to be 25\%. Our choice is dictated by a trade-off between selecting the lowest possible level of intensity (to be more sensitive to the presence of wings) and keeping this level above the sensitivity limit. To optimize the signal-to-noise ratio, we smoothed the spectra to a resolution of 2.7~\kms\ before determining the FWZI, which made it possible to derive the FWZI for 70 sources out of 99 detected in SiO. No significant difference is found between the FWZI distributions of the sources with Lyman excess and those without, as indicated by the Kolmogorov-Smirnov test, which gives a probability of 91\% that the two distributions are equivalent.

While an analysis of the FWZI is useful to investigate possible differences between the two samples, it is not ideal to establish the presence of line wings. For this purpose we developed a different method that we describe below. For a perfectly Gaussian profile, the ratio between the FWZI (defined as above) and the full width at half maximum (FWHM) is fixed and equal to $\sqrt{\ln25/\ln2}\simeq1.41$. If, instead, broad wings are present, the ratio FWZI/FWHM must be significantly greater. In an attempt to determine the presence of line wings in our sources in an objective way, we computed FWZI/FWHM for the SiO(2--1) line. We note that we were unable to use the \HCOp(1--0) line, which has also been detected in outflows (e.g., \citealt{sep10}), because the line shape is often asymmetric and makes it impossible to obtain a reliable measurement of the FWHM.

Figure~\ref{ffws} shows the distributions of the ratio FWZI/FWHM for sources with and without Lyman excess. Although some evidence of non-Gaussian profiles is seen (i.e., several sources with FWZI/FWHM$>$1.41, suggesting the presence of outflows), the important point is that no remarkable difference is found between the two distributions, as confirmed by the Kolmogorov-Smirnov statistical test, which gives a relatively high probability (28\%) that the two samples have the same intrinsic distribution. It is also worth noting that sources with prominent wings (e.g., with FWZI/FWHM$>$2), correspond to 28$\pm$12\% of the Lyman-excess sample and 34$\pm$6\% of the other sample: these percentages are comparable, within the uncertainties, and therefore support the similarity between the two types of objects.

In conclusion, molecular outflows do not appear to be preferentially associated with Lyman-excess sources, and neither the flashlight effect, nor any other outflow-related phenomenon can therefore be a viable explanation.

\subsection{Infall}

To determine the presence of infall, we adopted the approach of \citet{mardo}. Their method is based on the fact that an optically thick line of a molecule tracing infall, such as \HCOp, has an asymmetric profile caused by redshifted self-absorption. In contrast, a line of the isotopolog, \HCOpI, being optically thin, has a Gaussian profile. Therefore, by comparing the peak velocities of the two isotopologs, it is possible to reveal the presence of infall. In practice, \citet{mardo} defined the parameter
\begin{equation}
\delta V = (V_{\rm thick}-V_{\rm thin})/\Delta V_{\rm thin},
\end{equation}
where \Vthick\ is the peak velocity of the \HCOp(1--0) line and \Vthin\ and \DVthin\ are, respectively, the peak velocity and FWHM of the \HCOpI(1--0) line. Infall and expansion correspond to \dv$<$0 and \dv$>$0, respectively.

\begin{figure}
\centering
\resizebox{7.0cm}{!}{\includegraphics[angle=0]{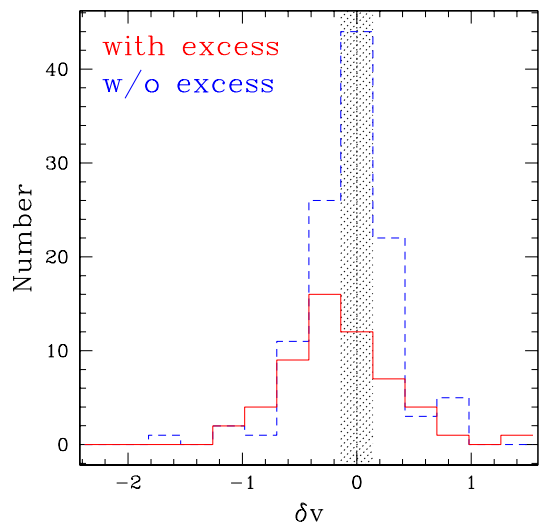}}
\caption{Distributions of the \citet{mardo} parameter, \dv, for sources with (red solid histogram) and without (blue dashed histogram) Lyman excess. The hashed area indicates the range inside which $|\delta V|<3\sigma$. Note how the solid distribution is skewed towards negative values of \dv.}
\label{fhdv}
\end{figure}

We computed \dv\ for 171 of our sources, 56 of these with Lyman excess. The remaining 29 objects were not detected in the \HCOpI\ line or presented complex spectra made of multiple components and/or were affected by deep absorption features. The distributions of \dv\ for the two samples are shown in Fig.~\ref{fhdv}. The step of the histograms has been taken equal to the mean 6$\sigma$ uncertainty on the \dv\ parameter. The distribution of the Lyman-excess sources is skewed toward negative values of \dv, whereas the other appears centered on \dv=0. The Kolmogorov-Smirnov test gives a probability of only 3\% that the two distributions are equivalent.

To further support that the Lyman-excess sample is dominated by infalling sources, we again followed \citet{mardo}, who defined the parameter
\begin{equation}
E=(N_{\rm infall}-N_{\rm expansion})/N_{\rm total}   \label{eqe}
,\end{equation}
where $N_{\rm infall}$ is the number of sources with $\delta V<-3\,\sigma$ ($1\,\sigma$ RMS in $\delta V$ is $\sim$0.05), $N_{\rm expansion}$ those with $\delta V>+3\,\sigma$, and $N_{\rm total}$=171 the number of all sources. Positive values of $E$ denote samples characterized by infalling objects, while negative values are associated with outflows. For the Lyman-excess sources we find $E=0.32\pm0.09$, whereas the other sample has $E=0.10\pm0.06$, consistent with the Lyman-excess sources being significantly more associated with infall than the others. In Table~\ref{tdet} we provide a summary of the information on the presence of infall in our sources.

Based on the above, we conclude that our findings are consistent with the accretion scenario proposed to explain the Lyman excess. Here a caveat is in order. Our \HCOp\ measurements are sensitive to large-scale ($\la$1~pc) infall and not to accretion onto the star or a putative circumstellar disk. The presence of \textup{\textup{\textup{{\it \textup{infall}}}}} is thus not direct evidence of the existence of {\it \textup{accretion}}. However, we believe it plausible that at least part of the infalling material is bound to become focused on a very small region and there dissipate its energy in shocks.

%
\section{Discussion and conclusions\label{sdis}}

The existence of Lyman excess in young early-type stars may have important consequences not only on the structure and evolution of \HII\ regions, but also on our understanding of the high-mass star formation process itself. All theories appear to agree that even the most massive stars may form through accretion, despite the different mechanisms invoked by the different models, such as competitive accretion \citep{bonn} and monolithic collapse \citep{krum}. However, it is still unclear whether the star reaches the ZAMS at approximately its final mass, or if still grows after igniting hydrogen burning.

Assuming that each of the \HII\ region is associated with a single massive star, our findings appear to support the latter scenario, where accretion may continue through the ionized region enshrouding the star, consistent with the model by \citet{keto03, keto07}. We can also conclude that the material accreted in this phase is a non-negligible fraction of the final stellar mass. If one-third of the ultracompact \HII\ regions are in the accretion phase, this means that accretion continues during one-third of the ultracompact \HII\ region lifetime ($\sim$$10^5$~yr; \citealt{wc89b}), namely $\sim3\times10^4$~yr. At an accretion rate of $10^{-4}$--$10^{-3}~M_\odot$~yr$^{-1}$, typical of massive young stellar objects (see, e.g., \citealt{fazal} and references therein), this corresponds to a total accreted mass of 3--30~$M_\odot$, a significant fraction of the mass of an early-type star. All the above indirectly suggests that the accretion flow significantly deviates from spherical symmetry. High accretion rates should quench the formation of an \HII\ region \citep{yorke, walms, keto02}, but if accretion is focused through a circumstellar disk, for instance, the Lyman-continuum photons may escape along the disk axis and hence ionize the surrounding gas. This could explain the co-existence of accretion onto the star and formation of a compact \HII\ region in the same object.

The scenario depicted above assumes that our \HII\ regions are ionized by single stars. However, massive stars form in clusters (e.g., \citealt{tan}), and it is likely that multiple stars contribute to the ionization. Therefore, an alternative scenario is that the infalling material does not end up on the ionizing, ZAMS star, but on one or more stellar companions still in the main accretion phase. In this case, our findings would imply that the high-mass star formation process in the cluster does not come to an end once the most massive star has reached the ZAMS, but instead continues even after the formation of an \HII\ region. Whether splitting the accreting material among two or more stars will be efficient enough to produce the required excess of Lyman-continuum photons is a non-trivial question that requires a dedicated model to be answered.

While all the previous considerations are at present purely speculative, direct detection of the accretion flow both in the molecular gas and in the ionized gas in the stellar surroundings are needed to establish the origin of the Lyman excess on solid grounds and to decide whether accretion is focused onto a single star or distributed among multiple stellar companions. To shed light on these questions, observations with the Atacama Large Millimeter and submillimeter Array (ALMA) are in progress for a selected subsample of our targets.

\begin{acknowledgements}
It is a pleasure to thank the staff of the 30 m telescope for their valuable support during the observations. The research leading to these results has received funding from the European Commission Seventh Framework Programme (FP/2007-2013) under grant agreement N.~283393 (RadioNet3).
\end{acknowledgements}

\end{document}